\documentclass[twocolumn,showpacs,preprintnumbers,amsmath,amssymb]{revtex4}
\usepackage{dcolumn}
\usepackage{bm}
\usepackage{graphicx}
\begin{document}

\title{Controlling transition probability from matter-wave soliton to chaos}
\author{Qianquan Zhu,  Wenhua Hai$^*$, Shiguang Rong}
\affiliation{Key Laboratory of Low-dimensional Quantum Structures
and Quantum Control of Ministry of Education, and \\ Department of
physics, Hunan Normal University, Changsha 410081, China}
\altaffiliation{Corresponding Author: W. Hai}
\email{whhai2005@yahoo.com.cn}

\begin{abstract}
For a Bose-Einstein condensate loaded into a weak traveling optical
superlattice it is demonstrated that under a stochastic initial set
and in a given parameter region the solitonic chaos appears with a
certain probability. Effects of the lattice depths and wave vectors
on the chaos probability are investigated analytically and
numerically, and different chaotic regions associated with different
chaos probabilities are found. The results suggest a feasible method
for eliminating or strengthening chaos by modulating the moving
superlattice experimentally. \vspace{0.2cm}

\end{abstract}

\maketitle

Many phenomena observed in Bose-Einstein condensates (BECs) are well
modelled by nonlinear Schr\"{o}dinger Equation (NLSE), also known as
Gross-Pitaevskii equation (GPE), which includes many fantastic
nonlinear effects, such as chaos\cite{Abdullaev,Zhang},
soliton\cite{Abdullaev2,Mayteevarunyoo}, and so on. Chaotic soliton
behaviors, which are of particular interest, have been studied
theoretically in a NLSE with a periodic perturbation\cite{11,12,13}.
Lately, there are a few remarkable works on chaotic dynamics of
soliton in BEC systems, including the chaos and energy
exchange\cite{14}, the discrete soliton and chaotic dynamics in an
array of BECs\cite{15}, and the bright matter-wave soliton
collision\cite{16}. Recently, increasing interest is excited in a
BEC held in an optical superlattices for the
periodic\cite{super2,super3} and quasiperiodic\cite{quasi1,ir1,ir2}
cases which are in close analogies with the fields of supercrystals
and quasicrystals\cite{qc1}. The BECs interacting with a traveling
lattice have also been successfully treated experimentally
\cite{mov1,mov2} and theoretically\cite{gap,chongPRE,lifei} that
shows many fantastic results such as lensing effect\cite{mov2}, gap
soliton generation\cite{gap}, spatiotemporal
chaos\cite{chongPRE,lifei}, and so on. From the above analyzes we
get a physical motivation, namely using a moving optical
superlattices to study the transitions from soliton to chaos in BEC
matter.

It is well known that for stochastic initial and boundary conditions
and fixed parameters, chaos does not always appear in a chaotic
system\cite{chongPRE}, but with certain probability. Chaos
probability may play a significant role for the control of chaos.
Many works have focused on suppressing chaos which results in zero
chaos probability. Whereas, in some realistic applications such as
secure communication based on chaos\cite{chaos}, higher chaoticity
is desired\cite{Li}, which calls for higher chaos probability. In
this paper, we show that in a BEC system perturbed by a moving
optical superlattice consists of two lattices of different depths
and wave vectors, the superlattice can separate the chaotic region
into several parts with different chaos probabilities. Furthermore,
for a fixed first lattice the adjustment to the secondary lattice
could turn the chaos probability to zero or higher one. The results
suggest a feasible method for eliminating or strengthening chaos
experimentally. Such a method can be extended to the controls of the
spatial chaos with zero traveling velocity and the temporal chaos in
other systems.

Primary and secondary moving optical lattice compose the
superlattice\cite{movsuper} as the form
\begin{eqnarray}\label{vlat}
V_{op}(\zeta)=V_1\cos^2(k\zeta)+V_2 \cos^2(\gamma k\zeta+\phi),
\end{eqnarray}
where we refer to $V_1 \cos^2(k\zeta)$ as the primary lattice with
$V_1$ and $k$ corresponding to its depth and wave vector, and $V_2
\cos^2(\gamma k\zeta+\phi)$ as the secondary one with $V_2$ being
its depth, $\gamma$ the ratio of the two laser wave vectors and
$\phi$ the phase difference. The spatiotemporal variable
$\zeta=x+v_Lt$ contains the velocity of the traveling lattice
$v_L=\delta/(2k)$ with $\delta$ being the frequency difference
between the two counter-propagating laser beams producing the first
lattice. The laser frequencies and amplitudes can be controlled
independently by using the acousto-optic modulators\cite{mov1}.

\textit{Chaotic Solitons} For the quasi-one-dimensional (1D) BEC
held in superlattice potential $V_{op}(\zeta)$ the transverse wave
function is approximately in ground state of a harmonic oscillator
of frequency $\omega_r$, and the governing time-dependent GPE reads
\begin{eqnarray}\label{1}
&&i\frac{\partial\Psi}{\partial
t}=-\frac{\hbar^2}{2m}\frac{\partial^2\Psi}{\partial
x^2}+g_{1d}|\Psi|^{2}\Psi+V_{op}(\zeta)\Psi.
\end{eqnarray}
Here $\Psi$ and $m$ are the macroscopic quantum wave function and
mass, $g_{1d}=2\hbar\omega_r a_s$ characterizes the 1D interatomic
interaction strength\cite{g} with $a_s$ being the s-wave scattering
length. In order to get a simple description, we investigate the
traveling wave solution in the form\cite{chongPRE}
\begin{eqnarray}\label{2}
\Psi=R(\zeta)e^{i[\theta(\zeta)+\alpha x+\beta t]},
\end{eqnarray}
where $R(\zeta)$ and $\theta(\zeta)$ are real functions of $\zeta$,
$\alpha$ and $\beta$ represent two undetermined real constants.
Inserting Eq.(\ref{2}) into  Eq.(\ref{1}), we easily obtain two
coupled equations
\begin{eqnarray}\label{R}
\frac{d^2R}{d
\xi^2}&-&R\Big(\frac{d\theta}{d\xi}\Big)^2-(v+2\tilde{\alpha})R\frac{d\theta}{d\xi}-(\tilde{\beta}+\tilde{\alpha}^2)R-g_1R^3
\nonumber\\&=&[\tilde{V_1}\cos^{2}(k\xi)+\tilde{V_2}\cos^{2}(\gamma
k\xi+\phi)]R, \\
\label{3} R\frac{d^2\theta}{d\xi^2}&+&2\frac{d
R}{d\xi}\frac{d\theta}{d\xi}+(v+2\tilde{\alpha})\frac{d R}{d\xi}=0.
\end{eqnarray}
In the formulas we have used the dimensionless variables and
parameters $\xi=k_0\zeta,\ v=2m v_L/\hbar k_0,\
\tilde{\beta}=\hbar\beta/E_{r0},\ \tilde{\alpha}=\alpha/k_0,\
\tilde{V_1}=V_1/E_{r0},\ \tilde{V_2}=V_2/E_{r0},\
E_{r0}=\hbar^2k_0^2/(2m)$ and $g_1=4a_s/(k_0l_r^2)$ with $k_0$ being
the unit of wave vector and $l_r=\sqrt{\hbar/(m\omega_r)}$ the
radical length of harmonic oscillator. Integrating Eq.(\ref{3})
yields $d\theta/d\xi=C/R^2-( v/2+\tilde{\alpha})$, where $C$ is the
integration constant. Applying this to Eq.(\ref{3}), we arrive at
the decoupled equation
\begin{eqnarray}\label{5}
&&\frac{d^2R}{d \xi^2}-\frac{C^2}{R^3}+D R-g_1R^3\nonumber\\
&&=[\tilde{V_1}\cos^{2}(k\xi)+\tilde{V_2}\cos^{2}(\gamma
k\xi+\phi)]R
\end{eqnarray}
with $D=\frac14v^2+v\tilde{\alpha}-\tilde{\beta}$. It is very hard
to find the exact solution of this equation. However, when the
driving strengths are weak enough, we can treat the chaotic system
by the direct perturbation approach\cite{haiweirao} and
Melnikov-function method\cite{melnikov}.

For simplicity we consider the $C=0$ case which leads Eq. (6) to the
driven Duffing equation, whose chaotic features have been
extensively studied for the single lattice
case\cite{chongPRE,lifei}. In the double lattice case, we expand $R$
to the first order,
\begin{eqnarray}
R(\zeta)=R_0(\zeta)+R_1(\zeta),\ \  |R_0|\gg|R_1|,
\end{eqnarray}
and insert it into Eq.(6) with $C=0$, obtaining the zeroth-order and
first-order equations
\begin{eqnarray}\label{zeroe}
\frac{d^2R_0}{d \zeta^2}+DR_0-g_1R_0^3=0,
\end{eqnarray}
\begin{eqnarray}\label{firste}
&&\frac{d^2R_1}{d \xi^2}+DR_1-3g_1R_0^2R_1=\varepsilon(\xi), \\
\nonumber
&&\varepsilon(\xi)=[\tilde{V_1}\cos^{2}(k\xi)+\tilde{{V_2}}\cos^{2}(\gamma
k\xi+\phi)]R_0.
\end{eqnarray}
If the atom-atom interactions are attractive, the system has a
negative $s$-wave scattering length to make $g_1<0$ such that
Eq.(\ref{zeroe}) with a negative $D$ has the well known homoclinic
solution
\begin{eqnarray}\label{zeros}
&&R_0(\xi)=\sqrt{\frac{2D}{g_1}}\textrm{sech}[\sqrt{-D}(\xi+c_0)],\\
\nonumber &&c_0=\frac{1}{\sqrt{-D}}\textrm{Ar
sech}\Big[\sqrt{\frac{g_1}{2D}}R_0(\xi_0)\Big]-\xi_0.
\end{eqnarray}
Here $\xi_0=k_0(x_0+v_L t_0)$ is the combination of the initial time
$t_0$ and boundary coordinate $x_0$, $c_0$ denotes an integration
constant determined by the boundary and initial conditions.
Obviously, the zeroth-order number density $R_0^2(\xi)$ is just a
bright soliton solution. Applying Eq.(\ref{zeros}) to
Eq.(\ref{firste}), we construct the general solution of
Eq.(\ref{firste}) in the integral form\cite{haiweirao}
\begin{eqnarray}\label{firsts}
R_1=h\int^\xi_P f\ \varepsilon(\xi)\ d\xi-f\int^\xi_Q h\
\varepsilon(\xi)\ d\xi,
\end{eqnarray}
where $P$ and $Q$ are the integration constants, $f=dR_0/d\xi$ and
$h=f\int f^{-2}d\xi$ are two linearly independent solutions of
Eq.(\ref{firste}) for $\varepsilon(\xi)=0$,
\begin{eqnarray}\label{fh}
f&=&\sqrt{\frac{2}{-g_1}}D\textrm{sech}[\sqrt{-D}(\xi+c_0)]\tanh[\sqrt{-D}(\xi+c_0)],
\nonumber\\
h&=&
-\frac{\sqrt{-2g_1}}{8(-D)^\frac23}\textrm{sech}[\sqrt{-D}(\xi+c_0)]\tanh[\sqrt{-D}(\xi+c_0)]\nonumber\\&&
\times\Big\{6\sqrt{-D}(\xi+c_0)-4\coth[\sqrt{-D}(\xi+c_0)]\nonumber\\&&+\sinh[2\sqrt{-D}(\xi+c_0)]\Big\}.
\end{eqnarray}
Generally, the corrected solution (\ref{firsts}) is unbounded,
because of the unboundedness of $h$ at $\xi=\infty$. However, using
the l'H\"{o}pital rule, we can easily obtain the sufficient and
necessary boundedness condition\cite{haiweirao}
\begin{eqnarray}\label{fd}
I_{\pm}=\lim_{\xi\rightarrow\pm\infty}\int^\xi_P f\
\varepsilon(\xi)\ d\xi=0.
\end{eqnarray}
It is worth noting that the relation between the Melnikov function
$M(c_0)$ and Eq.(\ref{fd}) is
$M(c_0)=I_+-I_-=\int^{+\infty}_{-\infty}f\ \varepsilon(\xi)\ d\xi$.
Combining $\varepsilon(\xi)$ in Eq.(\ref{firste}) with $R_0$ in
Eq.(\ref{zeros}) and $f$ in Eq.(\ref{fh}) to the integrand, we yield
the well-known Melnikov chaos criterion\cite{melnikov}
\begin{eqnarray}\label{qy}
&&M(c_0)=\frac{2k^2\pi}{g_1}\Big[\tilde{V_1}\textrm{csch}\Big(\frac{k\pi}{\sqrt{-D}}\Big)\sin(2c_0k)\nonumber\\
&&+\tilde{V_2}\gamma^2\textrm{csch}\Big(\frac{\gamma
k\pi}{\sqrt{-D}}\Big)\sin(2c_0\gamma k-2\phi)\Big]=0,
\end{eqnarray}
which indicates the existence of chaos for some $c_0$ values. Under
the conditions of Eqs.(\ref{fd}) and (\ref{qy}) we can call
Eq.(\ref{firsts}) the ``chaotic solution". Thus inserting Eqs. (10)
and (11) into Eq. (7) produces the chaotic bright soliton solution
which is the superposition between the soliton and chaotic states
and propagates with the velocity of traveling superlattice.

\textit{Chaos Probabilities and Regions}  The Melnikov function is a
periodic function of $c_0$ for the fixed parameters and rational
number $\gamma$, so only the discrete zero points $c_0=c_{0i}$ for
$i=1,2,\cdots$ satisfy the chaos criterion $M(c_{0i})=0$.
Nevertheless, $c_0$ is an integration constant depending on the
initial and boundary conditions and cannot be accurately set in
experiment. When the stochastic initial and boundary conditions are
applied to the numerical or experimental investigations, from Eq.
(10) we know that $c_{0}$ takes $c_{0i}$ values with only a certain
probability which is just the probability from soliton to chaos.
Defining this probability as the chaos probability $P$, it is clear
that the $P$ value should be proportional to the number $n$ of
$c_{0i}$ in one period of $M(c_0)$ and is always smaller than $1$.
Taking the chaos probability of single lattice case as the
referential one $P_0$, we have the relation $P=n P_0/2$ between $P$
and $n$, since $n$ is equal to $2$ in the single lattice case as in
Eq. (14) with $\tilde{V_2}=0$.
\begin{figure}[htp]
\center
\includegraphics[width=2in]{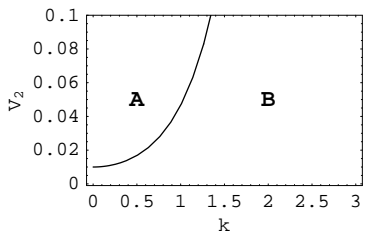}
\caption{\scriptsize {Plot of the chaotic regions of $\tilde{V_2}$
versus $k$ for the dimensionless parameters  $\gamma=2,\ D=-2$ and
$\tilde{V_1}=0.04$.}}
\end{figure}

Now let us see how the chaos probability depends on the parameter
regions. As a simple example we firstly consider the case $\phi=0$
and $\gamma=N\ge 2$ with $N$ integer. We rewrite Eq.(\ref{qy}) in
the form
\begin{eqnarray}\label{u1}
M(c_0)=\eta X_1(c_0)X_N(c_0)=0,
\end{eqnarray}
where $\eta=2k^2\pi/g_1$, $X_1(c_0)=\sin(2c_0k)$, and $X_N(c_0)=
\tilde{V_1}\textrm{csch}(k\pi/\sqrt{-D}) +F_N(c_0)$ with
$F_N(c_0)=\tilde{V_2}N^2\textrm{csch}(\frac{N
k\pi}{\sqrt{-D}})\frac{\sin(N2c_0k)}{\sin(2c_0k)}$. Clearly, the
$X_N(c_0)$ is a periodic function of $c_0$ but with different number
of zero points in different parameter regions, which can lead to
different number $n$ of $c_{0i}$ and different chaos probabilities.
This can be easily evidenced in the case $N=2$. The superlattice of
Eq.(1) with $\gamma=2$ has be widely studied\cite{movsuper}. In such
a case, the $X_N(c_0)$ becomes
$X_2(c_0)=\tilde{V_1}\textrm{csch}(k\pi/\sqrt{-D})
+8\tilde{V_2}\textrm{csch}(2k\pi/\sqrt{-D})\cos(2c_0k)$ and $M(c_0)$
has period $\pi$. Obviously, two parameter regions with different
zero points of $X_2(c_0)$ exist and the boundary between them is
given by
\begin{eqnarray}\label{u1}
\tilde{V_2}=\tilde{V}_{2b}=\frac{\tilde{V_1}\textrm{csch}(k\pi/\sqrt{-D})}{8\textrm{csch}(2k\pi/\sqrt{-D})}
=\frac{\tilde{V_1}}{4}\textrm{cosh}\frac{k\pi}{\sqrt{-D}}.
\end{eqnarray}
As an instance from Eq. (16) with $D=-2, \tilde{V_1}=0.04$ the
boundary curve on the parameter plane ($k,\tilde{V_2}$) is plotted
as in Fig. 1. The region $A$ above the boundary curve is associated
with $\tilde{V_2}>\tilde{V}_{2b}$ and the region $B$ below the
boundary curve corresponds to $\tilde{V_2}<\tilde{V}_{2b}$.

In order to show the different numbers $n$ of zero-point $c_{0i}$ in
one period of $M(c_0)$ for the different parameter regions, we make
the sketch maps showing the zero points of functions $X_1(c_0)$,
$X_2(c_0)$ and $M(c_0)$ as in Fig. 2. For the single lattice case
with $\tilde{V_2}=0$ from Fig. 2(a) we see that $X_2(c_0)=$
constant, $M(c_0)$ and $X_1(c_0)$ have $n=2$ same zero points,
indicating the chaos probability $P=nP_0/2=P_0$. In Fig. 2(b) with
parameters of region $A$, we exhibit that both $X_1(c_0)$ and
$X_2(c_0)$ have two zero points in range $c_0\in[0,\pi)$
respectively, and all the zero points are not coincident. This means
$n=4$ and the chaos probability $P=nP_0/2=2P_0$ in region $A$. On
the boundary between regions $A$ and $B$, from Fig. 2(c) we can see
that two zero points coincide among the three zero points of
$X_1(c_0)$ and $X_2(c_0)$ in one period. Therefore, we have $n=2$
and $P=P_0$ on the boundary curve. The same chaos probability
appears in region $B$, which is illustrated by Fig. 2(d), where
$X_2(c_0)$ has no zero point and the number $n=2$ of zero points of
$M(c_0)$ agrees with that of $X_1(c_0)$.
\begin{figure}[htp]
\center
\includegraphics[width=1.5in]{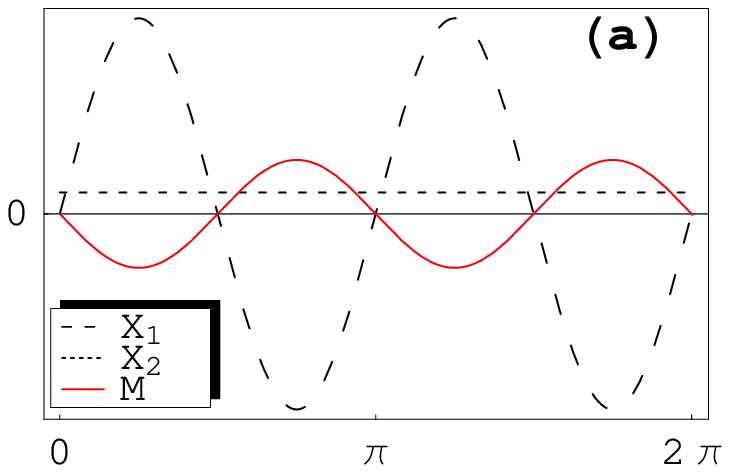}
\includegraphics[width=1.5in]{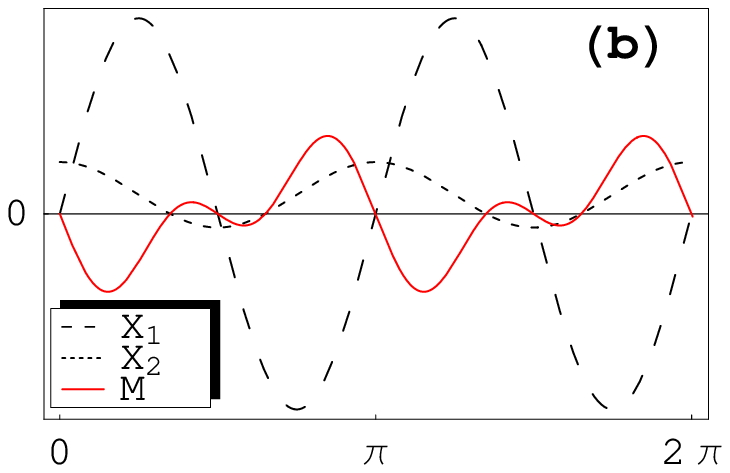}
\includegraphics[width=1.5in]{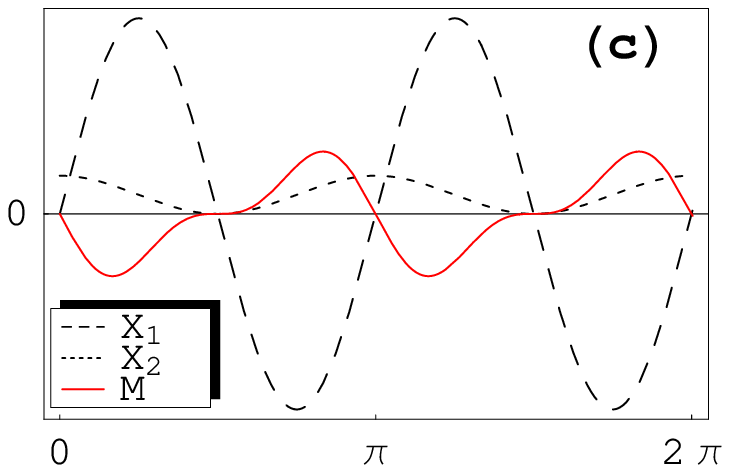}
\includegraphics[width=1.5in]{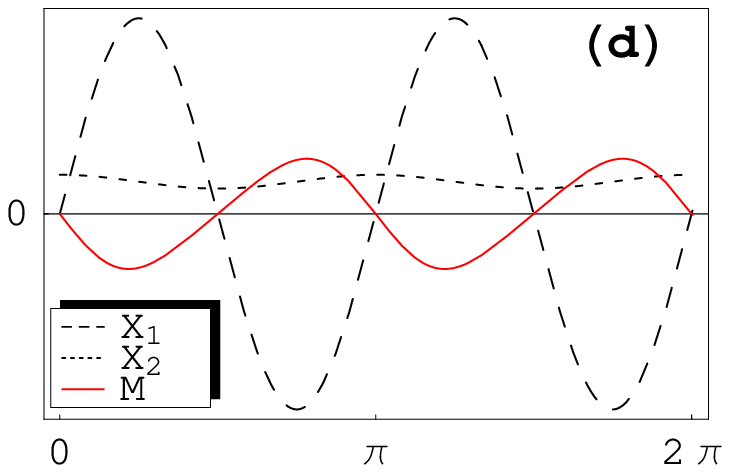}
\caption{\scriptsize {Sketch maps of functions $X_1(c_0)$ (dashed
curve), $X_2(c_0)$ (doted curve) and $M(c_0)$ (solid curve) versus
$c_0$ for the parameters $\gamma=2,\ D=-2,\ g_1=-0.5,\ k=1, \
\tilde{V_1}=0.04$ and (a) $\tilde{V_2}=0$, (b) $\tilde{V_2}=0.08
> \tilde{V}_{2b}$, (c) $\tilde{V_2}=0.01\cosh(\pi/\sqrt{2})
=0.0466\approx \tilde{V}_{2b}$, (d)
$\tilde{V_2}=0.01<\tilde{V}_{2b}$, where for each different function
the different amplitude scale is adopted for showing the zero
points.}}
\end{figure}

The above results can be classified as the two cases:

\bf Case 1. \rm In the double chaotic region $A$ with parameters
obeying $\tilde{V_2} > \tilde{V}_{2b}$, the chaos probability reads
$P=2P_0$;

\bf Case 2. \rm In the chaotic region $B$ and on the boundary curve
with parameters obeying $\tilde{V_2} \le \tilde{V}_{2b}$, the chaos
probability equates the referential one, $P=P_0$.

To numerically confirm the analytical results, we use the
MATHEMATICA code
\begin{eqnarray}
&&T=\pi,e[\{Rnew_,vnew_\}]:=\{R[T],v[T]\}/.Flatten \nonumber
\\ &&[NDSolve[\{R'[\xi]==v[\xi],v'[\xi] ==g_1R[\xi]^3-DR[\xi]\nonumber \\
&&+(V_1 Cos[k\xi]^2+V_2 Cos[\gamma
k\xi]^2)R[\xi],R[0]==Rnew,\nonumber
\\
&& v[0]==vnew\},\{R,v\},\{\xi,0,T\}]];Do[pic_i=ListPlot\nonumber
\\ && [Drop[Nestlist[e,\{Random[Real,\{-2.1,2.1\}],\nonumber \\ &&Random[Real,\{-2.2,2.2\}]\},3000],100],\{i,1,500\}]
\nonumber
\end{eqnarray}
to make two groups of Poincar\'{e} sections on the equivalent phase
space $(R, R_\xi)$ for the parameters used in Fig. 2(b) and Fig.
2(d) and the random initial conditions $\{R(\xi_0)\in[-2.1,2.1],\
R_\xi(\xi_0)\in[-2.2,2.2]\}$ associated with a suitable range of
$c_0$. Each of the groups contains five hundreds of Poincar\'{e}
sections. For the parameters in region $A$, $180$ chaotic attractors
are found, and the number of chaotic attractors is $85$ for the
parameters in region $B$. The numerical results show that in the
chaotic region $B$ the chaos probability reads $P=P_0=85/500=0.17$,
and in the double chaotic region $A$ the chaos probability becomes
$P=180/500=0.36\approx 2P_0$. They well agree with the analytical
assertion. The small difference exists between the analytical and
numerical results, because of that the strict chaos probability
requires more samplings of the Poincar\'{e} sections than the used
ones.

Then, we take $\phi=0$ and $\gamma =N=3,4,5$ into account and find
that for any $N$ the parameter space can be divided into several
parts with different chaos probabilities. The lowest chaos
probability is $P_0$ and the highest one arrives at $NP_0$.
Moreover, the number of chaotic regions increases with $N$.

\textit{Control of Chaos Probabilities} Controlling the chaos
probability to zero is important for some cases one requires
suppressing chaos. On the other hand, for the secure chaos-based
cryptosystems\cite{chaos,Li} one calls for higher chaotic
probability. Given the above-mentioned results, we can decrease or
raise the chaos probability by adjusting the parameters into
different chaotic regions. In order to suppress chaos, at first we
let the parameters enter the region of the lowest chaos probability
(e.g. the above region $B$), where only $X_1(c_0)=\sin(2c_0k)$ has
zero points. For such a parameter region if a chaotic state is
observed experimentally, which implies $2c_0k=i\pi$ for
$i=1,2,\cdots$ and the first term of Eq. (14) vanishing. Then we can
adopt the two different methods to eliminate chaos:

Method 1. We take $\phi=0$ and vary $\gamma$ from an integer to an
irrational number suddenly such that the second term in Eq. (14),
$\sin 2\gamma c_0k=\sin i \gamma \pi$, does not vanish and the
Melnikov chaos criterion $M(c_0)=0$ cannot be fulfilled, namely
$n=0$ and $P=nP_0/2=0$.

Method 2. Let $\gamma$ be equal to $2$ and suddenly change $\phi$
from zero to nonzero, say $\pi/4$, the nonzero Melnikov function can
also be obtained, that leads the chaos probability to zero.

The operation of method 1 can be performed by switching off the
second lattice with $\gamma=$ integer suddenly and switching on the
third lattice with $\gamma=$ irrational number simultaneously.
Similarly, we can complete the operation of method 2 by a new
lattice of phase difference $\pi/4$ with the second lattice. The
analysis offers an effective technique for suppressing
spatiotemporal chaos experimentally.

In conclusion, we have investigated a BEC system loaded into a weak
moving optical superlattice created by interference between two
lattices of different depths and wave vectors. The superlattice
separates the chaotic region into several parts with different chaos
probabilities, which are studied analytically and numerically.
Moreover, for a fixed first lattice, the modulation of the secondary
lattice can transform the chaos probability to zero or higher one.
Our results suggest a feasible method for suppressing or
strengthening chaos experimentally. When the zero traveling velocity
$v_L=0$ is set, the results fit the corresponding static BEC system.
The used method could also be easily extended to the controls of
other chaotic systems.

\begin{acknowledgments}
This work was supported by the National Natural Science Foundation
of China under Grant No. 10575034.
\end{acknowledgments}


\begin{references}
\bibitem{Abdullaev} F. Kh. Abdullaev, and R. A. Kraenkel, Phys. Rev. A \textbf{62}, 023613 (2000).
\bibitem{Zhang} C. Zhang, J. Liu, M. G. Raizen, and Q. Niu, Phys. Rev. Lett. \textbf{93}, 074101 (2004).
\bibitem{Abdullaev2} F Kh Abdullaev, and R Galimzyanov,  J. Phys. B
\textbf{36}, 1099 (2003).
\bibitem{Mayteevarunyoo} T. Mayteevarunyoo, B. A. Malomed, and M. Krairiksh, Phys. Rev. A \textbf{76}, 053612
(2007).
\bibitem{11} K. Nozaki, and N. Bekki, Phys. Rev. Lett. \textbf{50}, 1226 (1983).
\bibitem{12} H. T. Moon, and M. V. Goldman, Phys. Rev. Lett. \textbf{53}, 1821 (1984).
\bibitem{13} R. Scharf and A. R. Bishop, Phys. Rev. A \textbf{46}, R2973 (1992).
\bibitem{14} P. V. Elyutin, A. V. Buryak, V. V. Gubernov, R. A. Sammut, and I. N.
Towers, Phys. Rev. E \textbf{64}, 016607 (2001).
\bibitem{15} F. Kh. Abdullaev, E. N. Tsoy, B. A. Malomed, and R. A.
Kraenkel, Phys. Rev. A \textbf{68}, 053606 (2003).
\bibitem{16} A. D. Martin, C. S. Adams, and S. A. Gardiner, Phys. Rev. Lett. \textbf{98}, 020402 (2007).
\bibitem{super2} R. Roth and K. Burnett, J. Opt. B \textbf{5}, S50
(2003).
\bibitem{super3} C. Huang and W. Wu, Phys. Rev. A \textbf{72}, 065601
(2005).
\bibitem{quasi1} L. Sanchez-Palencia and L. Santos, Phys. Rev. A \textbf{72}, 053607 (2005).
\bibitem{ir1} A. A. Sukhorukov, Phys. Rev. Lett. \textbf{96}, 113902 (2006).
\bibitem{ir2} J. E. Lye, L. Fallani, C. Fort, V. Guarrera, M. Modugno, D. S. Wiersma, and M. Inguscio,
Phys. Rev. A \textbf{75}, 061603(R) (2007).
\bibitem{qc1} D. Shechtman, I. Blech, D. Gratias and J. W. Cahn, Phys. Rev. Lett. \textbf{53}, 1951
(1984).
\bibitem{mov1} J. Hecker Denschlag, J. E. Simsarian, H. H$\ddot{a}$fner, C. McKenzie,
A. Browaeys, D. Cho, K. Helmerson, S. L. Rolston, and W. D.
Phillips, J. Phys. B \textbf{35}, 3095 (2002).
\bibitem{mov2} L. Fallani1, F. S. Cataliotti1, J. Catani, C. Fort, M. Modugno, M. Zawada, and M.
Inguscio, Phys. Rev. Lett. \textbf{91}, 240405 (2003).
\bibitem{gap} B. Eiermann, Th. Anker, M. Albiez, M. Taglieber, P. Treutlein, K. P. Marzlin, and M. K.
Oberthaler, Phys. Rev. Lett. \textbf{92}, 230401 (2004).
\bibitem{chongPRE} G. Chong, W. Hai, and Q. Xie, Phys. Rev. E \textbf{70}, 036213
(2004).
\bibitem{lifei} F. Li, W. Shu, J. Jiang, H. Luo, and Z. Ren, Eur. Phys. J. D
\textbf{41}, 355(2007).
\bibitem{chaos} K. M. Cuomo and A. V. Oppenheim, Phys. Rev. Lett. \textbf{71},
65 (1993).
\bibitem{Li} X. Li, H. Zhang, Y. Xue, and G. Hu, Phys. Rev. E\textbf{71},
016216 (2005).
\bibitem{movsuper} P. J. Y. Louis, E. A. Ostrovskaya, and Y. S.
Kivshar, Phys. Rev. A \textbf{71}, 023612 (2005).
\bibitem{g} S. A. Gardiner, D. Jaksch, R. Dum, J. I. Cirac, and P.
Zoller, Phys. Rev. A \textbf{62}, 023612 (2000).
\bibitem{haiweirao} W. Hai, C. Lee, G. Chong and
L. Shi, Phys. Rev. E \textbf{66}, 026202(2002);  C. Lee, W. Hai, L.
Shi, X. Zhu and K. Gao, Phys. Rev. A\textbf{64}, 053604(2001); W.
Hai, X. Liu, J. Fang, X. Zhang, W. Huang, G. Chong, Phys. Lett. A
275, \textbf{54} (2000).
\bibitem{melnikov}  V. K. Melnikov, Trans. Moscow Math. Soc.
\textbf{12}, 1 (1963).

\end{references}
\end{document}